\documentclass[aps,prc,amsmath,preprint,superscriptaddress,showkeys,showpacs]{revtex4-1}
\usepackage{graphicx}
\usepackage{epsfig}
\newcommand{\beqy}{\begin{eqnarray}}
\newcommand{\eeqy}{\end{eqnarray}}

\begin{document}

\title{Neutron drip transition in accreting and nonaccreting neutron star crusts}
\author{N. Chamel}
\affiliation{Institut d'Astronomie et d'Astrophysique, CP-226, Universit\'e Libre de Bruxelles,
1050 Brussels, Belgium}
\author{A.~F. Fantina}
\affiliation{Institut d'Astronomie et d'Astrophysique, CP-226, Universit\'e Libre de Bruxelles,
1050 Brussels, Belgium}
\author{J.~L. Zdunik}
\affiliation{N. Copernicus Astronomical Center, Polish Academy of Sciences, Bartycka 18, PL-00-716, Warszawa, Poland}
\author{P. Haensel}
\affiliation{N. Copernicus Astronomical Center, Polish Academy of Sciences, Bartycka 18, PL-00-716, Warszawa, Poland}

\date{\today}

\begin{abstract}
The neutron-drip transition in the dense matter constituting the interior of neutron stars generally
refers to the appearance of unbound neutrons as the matter density reaches some threshold density
$\rho_\textrm{drip}$. This transition has been mainly studied under the cold catalyzed matter hypothesis.
However, this assumption is unrealistic for accreting neutron stars. After examining the physical
processes that are thought to be allowed in both accreting and nonaccreting neutron stars, suitable
conditions for the onset of neutron drip are derived and general analytical expressions for the neutron
drip density and pressure are obtained. Moreover, we show that the neutron-drip transition occurs at
lower density and pressure than those predicted within the mean-nucleus approximation. This transition
is studied numerically for various initial composition of the ashes from X-ray bursts and superbursts using 
microscopic nuclear mass models.
\end{abstract}

\keywords{neutron drip, electron capture, dense matter, neutron emission, neutron star}

\pacs{97.60.Jd, 26.60.Gj, 26.60.Kp, 23.40.-s}

\maketitle

\section{Introduction}
\label{intro}

Born in catastrophic gravitational core-collapse supernova explosions, neutron stars are the densest
stars known in the universe~\cite{hae07}. According to our current understanding, a neutron star
contains qualitatively distinct regions. A thin atmospheric plasma layer of light
elements (mainly hydrogen and helium) possibly surrounds a Coulomb liquid of electrons and ions. Below
these liquid surface layers, the matter consists of a solid crust made of a crystal lattice of fully
ionized atoms. With increasing density, nuclei become progressively more neutron rich due to electron
captures until neutrons start to drip out of nuclei at some threshold density $\rho_\textrm{drip}$ (we shall 
use the symbol $\rho$ to denote the mass-energy density). 
This so called neutron-drip transition marks the boundary between the outer and inner crusts. The crust
dissolves into an homogeneous liquid mixture at about half the density prevailing in heavy atomic nuclei.

The presence of a neutron liquid in the inner crust is expected to play a role in various
observed astrophysical phenomena such as pulsar sudden spin-ups (so
called glitches), and the thermal relaxation of transiently
accreting neutron stars (see, e.g., Ref.~\cite{lrr}). Determining
the onset of neutron drip in dense matter is therefore of utmost
importance for the modelisation of these phenomena. This transition
is generally found to occur at the density $\rho_\textrm{drip}\simeq
4.3\times 10^{11}$~g~cm$^{-3}$ in nonaccreting neutron stars (see,
e.g., Refs.~\cite{bps,pearson2011,wolf13,hemp13}). This density can
be shifted due to the accretion of matter from a companion star. For
instance, the neutron drip density was found to be given by
$\rho_\textrm{drip}\simeq 6.1\times 10^{11}$~g~cm$^{-3}$
($\rho_\textrm{drip}\simeq 7.8\times 10^{11}$~g~cm$^{-3}$) assuming
that the ashes of the X-ray bursts consist of pure $^{56}$Fe
($^{106}$Pd respectively)~\cite{hz1990a,hz2003}. 
On the other hand,
these calculations lead to a discontinuous change of the unbound
neutron density hence also of the neutron chemical potential at the transition. 
This means that the crust is not locally in ``chemical'' equilibrium: free neutrons 
will be subject  to ``chemical'' forces and will thereby diffuse until the equilibrium 
is reached.

In this paper, the condition for the onset of neutron drip is examined considering the instability of
dense matter against electron captures and neutron emission. Our model of dense matter is briefly presented
in Section~\ref{model}. The usual determination of the neutron-drip transition in nonaccreting neutron stars
is discussed in Section~\ref{misconceptions}, and some misconceptions are pointed out. In particular, we show
that the equilibrium nuclei at the onset of neutron drip are stable against neutron emission. The stability of
dense matter is studied in Section~\ref{stability}, and applications to neutron stars are discussed in
Section~\ref{applications}. The cases of nonaccreting and accreting neutron stars are considered separately.

\section{Model of dense matter}
\label{model}

We consider matter at densities high enough that atoms are fully ionized. We further assume that
the temperature $T$ is lower than the crystallization temperature $T_m$ so that atomic nuclei are
arranged in a regular crystal lattice.
For simplicity, we consider crystalline structures made of only
one type of ions  $^A_ZX$ with proton number $Z$ and mass number $A$. In this case,
$T_m$ is given by (see, e.g., Ref.~\cite{hae07})
\begin{equation}
T_m=\frac{e^2}{a_e k_\text{B} \Gamma_m}Z^{5/3}\, ,
\end{equation}
$e$ being the elementary electric charge, $a_e=(3/(4\pi n_e))^{1/3}$ the electron-sphere radius, $n_e$ the electron number density,
$k_\text{B}$ Boltzmann's constant, and $\Gamma_m\simeq 175$ the Coulomb coupling parameter at
melting. Since $T_m$ is generally much lower than the electron Fermi temperature defined by
\begin{equation}
 T_\text{F}=\frac{\mu_e-m_e c^2}{k_\text{B}}\, ,
\end{equation}
where $\mu_e$ is the free electron chemical potential, and $m_e$ the electron mass, electrons
are highly degenerate. Charge screening effects are quite negligible so that electrons will be
assumed to be uniformly distributed. We will also ignore the small contribution due to electron exchange
and correlation to the energy density and pressure.
The expression of the electron energy density $\mathcal{E}_e$ and pressure $P_e$ 
in the limit $T\ll T_\text{F}$ can be found in Chap. 2 of Ref.~\cite{hae07}.
The free electron chemical
potential is given by
\begin{equation}
\mu_e=m_e c^2 \sqrt{1+x_r^2}\, ,
\end{equation}
where $x_r=\lambda_e (3\pi^2 n_e)^{1/3}$ is a dimensionless relativity parameter and $\lambda_e=\hbar/(m_e c)$ is the
electron Compton wavelength.
For point-like ions embedded in a uniform electron gas with number density $n_e$,
the lattice pressure is simply given by
\begin{equation}\label{eq:PL}
 P_L=\frac{\mathcal{E}_L}{3}\, ,
\end{equation}
where the lattice energy density $\mathcal{E}_L$ is given by (see e.g. Chap. 2 in Ref.~\cite{hae07})
\begin{equation}\label{eq:EL}
\mathcal{E}_L  =C e^2 n_e^{4/3} Z^{2/3}\, ,
\end{equation}
and the crystal structure constant is approximately given by 
$C\approx -1.44$ \cite{baiko01}
(we have omitted here the
small contribution due to quantum zero-point motion of ions about their equilibrium position).
In the ultrarelativistic regime $x_r\gg 1$, the free electron chemical potential and the total pressure $P=P_e+P_L$ 
can be approximately expressed as
\begin{equation}\label{eq:mue-ultra-rel}
\mu_e\approx \hbar c (3\pi^2 n_e)^{1/3}\, ,
\end{equation}
\begin{equation}\label{eq:P-ultra-rel}
P\approx \frac{\mu_e^4}{12 \pi^2 (\hbar c)^3}\left(1+ \frac{4 C \alpha Z^{2/3}}{(81\pi^2)^{1/3}} \right) \, ,
\end{equation}
where $\alpha=e^2/\hbar c$ is the fine structure constant.

\section{Misconceptions about the neutron drip transition in neutron stars}
\label{misconceptions}

According to the cold catalyzed matter hypothesis~\cite{hw58,htww65} that will be further discussed in Section \ref{applications}, 
the interior of a mature neutron star is in a state of full thermodynamic equilibrium with respect to all kinds 
of nuclear and electroweak processes at zero temperature. 
The properties of any layer of the outer crust at some pressure $P$ can thus be determined by minimizing the Gibbs free energy per 
nucleon $g$, defined by
\begin{equation}
\label{eq:gibbs-general}
g=\frac{\mathcal{E}+P}{n}\, ,
\end{equation}
where $n$ is the average nucleon number density, and $\mathcal{E}$ is the average energy density. In order to
determine the onset of neutron drip, we have to express the Gibbs free energy per nucleon for nuclei coexisting
with free neutrons and electrons.
In the shallowest region of the inner crust which we are interested in, the interactions between nuclei and free neutrons
can be neglected, as shown in Refs. \cite{cha06,cha07}. The average energy density is thus given by
\begin{equation}\label{eq:energy}
\mathcal{E}=n_X M(A,Z)c^2+\mathcal{E}_e+\mathcal{E}_L+  \mathcal{E}_n- n_e m_e c^2 \, ,
\end{equation}
where $n_X$ is the number densities of nuclei $^A_ZX$, $M(A,Z)$ their mass (including the rest mass of $A$ nucleons and
$Z$ electrons)  and $\mathcal{E}_n$ the average energy density of the neutron gas. 
The reason for including the electron rest mass in $M(A,Z)$ is that experimental \emph{atomic} masses are generally 
tabulated rather than \emph{nuclear} masses. The last term in Eq.(\ref{eq:energy}) is introduced to avoid double counting. 
As pointed in Ref. \cite{cch05}, the concept of ''free'' neutrons is ambiguous. In the following we shall consider that a neutron
is free if it is unbound in the quantum mechanical sense: its single particle energy in a self-consistent mean-field treatment lies
above the maximum value of the mean potential.
Using the thermodynamic identities $\mathcal{E}_e+P_e=n_e\mu_e$ and $\mathcal{E}_n+P_n=n_n\mu_n$
($P_n$ is the neutron contribution to the pressure, $n_n$ the neutron density and $\mu_n$ is the neutron chemical potential),
the Gibbs free energy per nucleon can be written as
\begin{eqnarray}\label{eq:gibbs-mean-nucleus}
g=\frac{n_X}{n} M(A,Z)c^2+\frac{n_e}{n}\biggl[\mu_e-m_e c^2+\frac{4}{3}\frac{\mathcal{E}_L}{n_e}\biggr] + \frac{n_n}{n} \mu_n\, .
\end{eqnarray}
Let us note that the average nucleon number density is given by
\begin{equation}\label{eq:nbaryon}
n=A n_X+n_n\, .
\end{equation}
Moreover, the electric charge neutrality yields
\begin{equation}
n_e=Z n_X\, .
\end{equation}
The Gibbs free energy per nucleon can finally be expressed as
\begin{eqnarray}\label{eq:gibbs-mean-nucleus2}
g(A,Z,n_e,y_n)=(1-y_n)\frac{M(A,Z)c^2}{A}+\frac{Z}{A}(1-y_n)\biggl[\mu_e-m_e c^2+\frac{4}{3}\frac{\mathcal{E}_L}{n_e}\biggr] + y_n \mu_n\, ,
\end{eqnarray}
where $y_n=n_n/n$ denotes the free neutron fraction.
In any layer of the outer crust, we have $y_n=0$ and $P=P_e+P_L$. Neutrons
start to drip out of nuclei at some pressure $P_\textrm{drip}=P_e(n_e^\textrm{drip})+P_L(n_e^\textrm{drip},Z)$ whenever
\begin{equation}
g(A, Z, P_\textrm{drip}, d y_n) = g(A, Z, P_\textrm{drip}, y_n=0)\, ,
\end{equation}
or equivalently
\begin{equation}\label{eq:n-drip-condition-gibbs}
\frac{\partial g}{\partial y_n}\biggr\vert_{Z,A,n_e^\textrm{drip}}=0\, ,
\end{equation}
where the partial derivative is evaluated for $y_n=0$. This latter condition leads to
\begin{equation}\label{eq:mun-drip}
 \mu_n(n_n=0)=g(A,Z,n_e^\textrm{drip},y_n=0)=\frac{M(A,Z)c^2}{A}+\frac{Z}{A}\biggl[\mu_e(n_e^\textrm{drip})-m_e c^2+\frac{4}{3}\frac{\mathcal{E}_L(n_e^\textrm{drip},Z)}{n_e^\textrm{drip}}\biggr]\, .
\end{equation}
At the neutron drip threshold, we have $\mu_n\approx m_n c^2$ (we neglect the small correction of the order of a few tens of keV 
due to neutron-band structure effects~\cite{cha06,cha07}).
Equation~(\ref{eq:mun-drip}) can thus be expressed as
\begin{equation}\label{eq:n-drip-mue}
 \mu_e(n_e^\textrm{drip}) + \frac{4}{3}C e^2 (n_e^{\textrm{drip}})^{1/3} Z^{2/3} =  \mu_e^{\textrm{drip}}\, ,
\end{equation}
where
\begin{equation}\label{eq:muedrip}
\mu_e^{\textrm{drip}}(A,Z)\equiv \frac{-M(A,Z)c^2+A m_n c^2}{Z} +m_e c^2 \, .
\end{equation}

Let us note that the equilibrium nucleus $^A_ZX$ at $P=P_\textrm{drip}$ must be such as to minimize the Gibbs free energy per nucleon. In other
words, we must have
\begin{equation}
g(A,Z,P_\textrm{drip},y_n=0)\leq g(A^\prime,Z^\prime,P_\textrm{drip},y_n=0)\, ,
\end{equation}
for any values of $A^\prime$ and $Z^\prime$. In particular, setting $Z^\prime=Z$ and $A^\prime=A-\Delta N$ 
using Eqs.~(\ref{eq:gibbs-mean-nucleus2}), (\ref{eq:mun-drip}), (\ref{eq:n-drip-mue}) and (\ref{eq:muedrip}),
we obtain
\begin{equation}\label{eq:cond-n-emission}
 M(A,Z)-M(A-\Delta N,Z)\leq \Delta N m_n\, .
\end{equation}
This shows that the nucleus $^A_ZX$ is actually stable against neutron emission
\begin{equation}\label{eq:n-emission}
^A_ZX \rightarrow ^{A-\Delta N}_{Z}X+ \Delta N n\, ,
\end{equation}
as previously noticed in Ref.~\cite{hz1989}.
As can be seen in Table~\ref{tab1}, the equilibrium nuclei at $P=P_{\rm drip}$ as predicted by various microscopic nuclear mass models
are indeed stable against neutron emission even though they lie beyond the neutron-drip line in the chart of
nuclides. Let us recall that this line is generally defined at each value of the proton number $Z$ by the lightest isotope for which
the neutron separation energy $S_n$, defined by
\begin{equation}
S_n(A,Z)\equiv M(A-1,Z)c^2-M(A,Z)c^2+m_nc^2\, ,
\end{equation}
is negative, i.e., the lightest isotope that is unstable with respect to neutron emission. Because of pairing and shell effects,
many nuclei beyond the neutron-rich side of the neutron drip line are actually stable.

\begin{table}
\centering
\caption{Equilibrium nucleus
at the bottom of the outer crust of nonaccreting neutron stars, as predicted by different microscopic nuclear mass
models~\cite{pearson2011}. The corresponding isotope
at the neutron-drip line in the nuclear chart is indicated. The numbers in
parenthesis are the neutron separation energies in MeV. The last two columns contain the neutron-drip density and pressure. }\smallskip
\label{tab1}
\begin{tabular}{ccccc}
\hline \noalign {\smallskip}
 & outer crust & drip line & $\rho_\textrm{drip}$ (g~cm$^{-3}$) & $P_\textrm{drip}$ (dyn~cm$^{-2}$)\\
\hline \noalign {\smallskip}
HFB-19 & $^{126}$Sr (0.73) & $^{121}$Sr (-0.62) & $4.40\times 10^{11}$ & $7.91\times 10^{29}$ \\
HFB-20 & $^{126}$Sr (0.48) & $^{121}$Sr (-0.71) & $4.39\times 10^{11}$ & $7.89\times 10^{29}$ \\
HFB-21 & $^{124}$Sr (0.83) & $^{121}$Sr (-0.33) & $4.30\times 10^{11}$ & $7.84\times 10^{29}$ \\
\hline
\end{tabular}
\end{table}

The fact that the equilibrium nucleus at the bottom of the outer crust is stable against neutron emission~(\ref{eq:n-emission})
has been often overlooked. The reason may be traced back to the use of semi-empirical mass formulae in the early studies of neutron-star crusts.
For instance, in their seminal work Harrison and Wheeler~\cite{hw58,htww65} (the same treatment can also be found in the
standard textbook from Shapiro and Teukolsky~\cite{shapiro1983}) considered that at high enough density
the nucleus $^A_ZX$ becomes unstable against neutron emission~(\ref{eq:n-emission}) with $\Delta N=1$, at which point each nucleus $^A_ZX$ is in
equilibrium with its isotope $^{A-1}_ZX$ and one free neutron, as embedded in Eq.(297) of Ref.~\cite{htww65}. In other words, they
considered that at the neutron drip transition
\begin{equation}\label{eq:n-emission-condition}
 M(A,Z)-M(A-1,Z)\geq m_n\, .
\end{equation}
Using a semi-empirical mass formula, they actually replaced this condition by its continuum version
\begin{equation}\label{eq:n-emission-condition-continuum}
 \frac{\partial M(A,Z)}{\partial A}\biggr\vert_Z \geq m_n\, .
\end{equation}
The equilibrium nucleus Harrison and Wheeler found at the interface between the outer and inner crusts is
$Z\simeq 39$ and $A\simeq 122$ (note from Table~\ref{tab1} that this is very close to the composition obtained
from recent microscopic nuclear mass models). It turns out that this
nucleus does not satisfy the inequality (\ref{eq:n-emission-condition}), i.e. it is stable against neutron emission (see Appendix). This stems from the
fact that the violation of the condition (\ref{eq:n-emission-condition}) corresponds to the emission of a huge number of neutrons,
one neutron for each nucleus present in the crustal layer of consideration. 
This cannot represent the physical processes associated with the
neutron-drip transition in neutron star crusts since this would lead to an unphysical discontinuous change of the neutron chemical potential.
On the contrary,
this transition can be viewed as the appearance of one free neutron in the whole crustal layer so that the neutron
fraction tends asymptotically to zero in the thermodynamic limit, and the neutron chemical potential thereby varies continuously. 
It is as if each nucleus were unstable against the emission
of an infinitesimally small fraction $\epsilon$ of a neutron
\begin{equation}\label{eq:n-drip}
^A_ZX \rightarrow ^{A-\epsilon}_{Z}X+ \epsilon\, n\, .
\end{equation}
Equation~(\ref{eq:n-emission-condition}) should thus be replaced by
\begin{equation}\label{eq:n-drip-condition-mass}
 M(A,Z)-M(A-\epsilon,Z)\geq \epsilon\, m_n\, ,
\end{equation}
which coincides with Eq.~(\ref{eq:n-emission-condition-continuum}) in the asymptotic limit $\epsilon\rightarrow 0$.
It can be shown that this continuum approximation leads to Eqs.~(\ref{eq:n-drip-mue}) and (\ref{eq:muedrip})
(see, e.g., Ref.~\cite{shapiro1983}).

Although the previous considerations shed some light on the neutron-drip transition in neutron star crusts, they still leave open the
question of the actual physical processes that can be collectively summarized by Eq.~(\ref{eq:n-drip}).
This question is of utmost importance for accreting neutron stars, whose crusts
are generally not in full thermodynamic equilibrium.

\section{Stability of dense matter against electron capture and neutron emission}
\label{stability}

\subsection{Mean-nucleus approximation}

With increasing density, matter becomes progressively more neutron rich due to  the capture of electrons, whereby the nucleus $^A_ZX$ transforms 
into a nucleus $^A_{Z-\Delta Z}Y$ with proton number $Z-\Delta Z$ and mass number $A$ with the emission of $\Delta Z$ electron neutrino $\nu_e$~:
\begin{equation}\label{eq:e-capture}
^A_ZX+\Delta Z e^- \rightarrow ^A_{Z-\Delta Z}Y+\Delta Z \nu_e\, .
\end{equation}
The nucleus $^A_ZX$ will be stable against such a process at some given pressure $P$
provided the corresponding Gibbs free energy per nucleon is lower than that of the daughter
nucleus $^A_{Z-\Delta Z}Y$. The expression of the Gibbs free energies per nucleon before
and after the electron capture can be obtained from Eq.~(\ref{eq:gibbs-mean-nucleus}).
After the electron capture, the average baryon number density will be given by
\begin{equation}
n^+=A n_Y\, ,
\end{equation}
where $n_Y$ denotes the density of nuclei $^{A}_{Z-\Delta Z}Y$.
Moreover, the electric charge neutrality requires
\begin{equation}\label{eq:ne+ny}
n_e^+=(Z-\Delta Z)n_Y=\frac{Z-\Delta Z}{A} n^+\, .
\end{equation}
The Gibbs free energy per nucleon after the electron capture can thus be finally expressed as
\begin{eqnarray}\label{eq:gibbs-e}
g(A,Z,\Delta Z, n_e^+)=&&\frac{M(A,Z-\Delta Z)c^2}{A}+\frac{Z-\Delta Z}{A}\biggl[\mu_e  + \frac{4}{3}\frac{\mathcal{E}_L(n_e^+,Z-\Delta Z)}{n_e^+} -m_e c^2 \biggr] \, .
\end{eqnarray}
Similarly, the Gibbs free energy per nucleon before the electron capture is simply given by
$g(A,Z,\Delta Z=0,n_e^-)$, with
\begin{equation}\label{eq:ne-ny}
n_e^-=Z n_X = \frac{Z}{A} n^-\, ,
\end{equation}
where $n^-$ denotes the average baryon density before the capture.
The electron densities $n_e^-$ and $n_e^+$ are not exactly the same because the
pressure $P$ has to remain constant during the process. Before the capture, the pressure
can be written as
\begin{equation}\label{eq:e-capture-P-}
P= P_e(n_e^-)+P_L(n_e^-,Z)\, .
\end{equation}
After the capture, the pressure can be expressed as
\begin{equation}\label{eq:e-capture-P+}
P=P_e(n_e^+)+P_L(n_e^+,Z-\Delta Z)\, .
\end{equation}
The lattice pressure is very small, $P_L\ll P_e$, so that $n_e^-\approx n_e^+$. Let us 
write $n_e^+ = n_e + \delta n_e$, where $n_e\equiv n_e^-$.  
Solving Eqs.~(\ref{eq:e-capture-P-}) and (\ref{eq:e-capture-P+}) to first order in $\delta n_e$ 
yields 
\begin{equation}\label{eq:deltane+}
 \delta n_e = \left[P_L(n_e,Z)-P_L(n_e,Z-\Delta Z)\right] \left(\frac{d P_e}{d n_e}\right)^{-1}<0\, .
\end{equation}
Considering ultrarelativistic electrons and using Eqs.~(\ref{eq:mue-ultra-rel}) and (\ref{eq:P-ultra-rel}), we find to first order in $\alpha$
\begin{equation}\label{eq:deltane+-rel}
 |\delta n_e| \approx -\frac{C\alpha}{(3\pi^2)^{1/3}} \biggl[Z^{2/3}-(Z-\Delta Z)^{2/3}\biggr] n_e \ll n_e\, .
\end{equation}
Whereas the electron density varies almost continuously $n_e^-\approx n_e^+$, the electron capture will be
accompanied by a discontinuous change of the baryon density given by
\begin{equation}
\frac{n^+-n^-}{n^-}\approx  \frac{\Delta Z}{Z-\Delta Z}\, .
\end{equation}

Expanding the stability condition $g(A,Z,\Delta Z=0,n_e^-)<g(A,Z,\Delta Z,n_e^+)$ to first order in $\alpha$ using 
Eq.~(\ref{eq:deltane+}) leads to
\begin{equation}\label{eq:e-capture+gibbs-approx-mean-nucleus}
 \mu_e + C e^2 n_e^{1/3}\biggl[ \frac{Z^{5/3}-(Z-\Delta Z)^{5/3}}{\Delta Z} + \frac{1}{3} Z^{2/3}\biggr] <  \mu_e^{\beta}\, ,
\end{equation}
where
\begin{equation}\label{eq:muebeta}
\mu_e^{\beta}(A,Z)\equiv \frac{M(A,Z-\Delta Z)c^2-M(A,Z)c^2}{\Delta Z} +m_e c^2 \, .
\end{equation}

With further compression of matter, the nucleus $^A_ZX$ may become unstable against the capture of electrons accompanied by the emission of free neutrons.
Let us consider the general case whereby the nucleus $^A_ZX$ transforms into a nucleus $^{A-\Delta N}_{Z-\Delta Z}Y$ with proton number $Z-\Delta Z$ and
mass number $A-\Delta N$ by capturing $\Delta Z$ electrons with the emission of $\Delta N$ neutrons $n$ and $\Delta Z$ electron neutrino $\nu_e$~:
\begin{equation}\label{eq:e-capture+n-emission}
^A_ZX+\Delta Z e^- \rightarrow ^{A-\Delta N}_{Z-\Delta Z}Y+\Delta N n+\Delta Z \nu_e\, .
\end{equation}
This process can occur at some pressure $P$ if it leads to a lower value for the Gibbs free
 energy per nucleon. After the electron capture, the average baryon number density is given by
\begin{equation}
n^+=(A-\Delta N) n_Y  + n_n\, ,
\end{equation}
where $n_Y$ is the density of nuclei $^{A-\Delta N}_{Z-\Delta Z}Y$.
Note that $\Delta N$ free neutrons are associated with each nucleus so that
\begin{equation}\label{eq:nn-ny}
 n_n=\Delta N n_Y\, .
\end{equation}
Therefore, the average baryon number density reduces to
\begin{equation}
n^+= A n_Y\, .
\end{equation}
The electric charge neutrality leads to
\begin{equation}
n_e^+=(Z-\Delta Z)n_Y\, .
\end{equation}
It follows from these equations that the neutron density can be equivalently expressed as
\begin{equation}\label{eq:nn-capture}
 n_n= \frac{\Delta N}{Z-\Delta Z} n_e^+ \, ,
\end{equation}
for $\Delta Z < Z$ and
\begin{equation}
n_n=n^+
\end{equation}
for $\Delta Z=Z$ (in this case we must obviously have $\Delta N=A$ and $n_e^+=0$). 
The Gibbs free energy per nucleon after the electron capture and the neutron emission can thus be written as
\begin{eqnarray}\label{eq:gibbs-e-mean-nucleus}
g(A,Z,\Delta Z, \Delta N, n_e^+)=&&\frac{M(A-\Delta N,Z-\Delta Z)c^2}{A}+ \frac{\Delta N}{A} \mu_n \\ \nonumber 
&+&\frac{Z-\Delta Z}{A}\biggl[\mu_e  + \frac{4}{3}\frac{\mathcal{E}_L(n_e^+,Z-\Delta Z)}{n_e^+} -m_e c^2
\biggr] 
\, .
\end{eqnarray}
The electron density $n_e^+$ can be obtained from the requirement that the process
occurs at a constant pressure $P$:
\begin{equation}\label{eq:e-capture-n-P-mean-nucleus}
P=P_e(n_e^+)+P_L(n_e^+,Z-\Delta Z)+P_n(n_n)\, .
\end{equation}
Setting $n_e^+\equiv n_e + \delta n_e$ with $n_e\equiv n_e^-$ in Eq.~(\ref{eq:e-capture-n-P-mean-nucleus}), and considering 
$\Delta Z<Z$, we find to first order in $\delta n_e$
\begin{equation}\label{eq:deltane+mean}
 \delta n_e = \left[P_L(n_e,Z)-P_L(n_e,Z-\Delta Z)-P_n\right] \left(\frac{dP_e}{dn_e}+\frac{\Delta N}{Z-\Delta Z}\frac{dP_n}{d n_n}\right)^{-1}\, ,
\end{equation}
where $dP_e/dn_e$ has to be evaluated at the density $n_e$, whereas $P_n$ and $dP_n/dn_n$ at the density $n_e \Delta N /( Z-\Delta Z)$. Typically 
$dP_n/dn_n \ll dP_e/dn_e$. We shall therefore neglect the derivative of the neutron pressure in Eq.~(\ref{eq:deltane+mean}). 
Expanding the stability condition $g(A,Z,\Delta Z=0,\Delta N=0, n_e^-)<g(A,Z,\Delta Z,\Delta N, n_e^+)$ to first order in $\alpha$ using 
Eqs.~(\ref{eq:gibbs-e-mean-nucleus}) and (\ref{eq:deltane+mean}), we find
\begin{eqnarray}\label{eq:e-capture+n-emission-gibbs-approx-mean-nucleus}
\mu_e &+& C e^2 n_e^{1/3}\biggl[\frac{Z^{5/3}-(Z-\Delta Z)^{5/3}}{\Delta Z} + \frac{1}{3} Z^{2/3} \biggr] -\left(1-\frac{Z}{\Delta Z}\right) \frac{P_n}{n_e}\\ \nonumber
&-& \frac{\Delta N}{\Delta Z}\left(\mu_n - m_n c^2\right) <  \mu_e^{\beta n} \, ,
\end{eqnarray}
where
\begin{equation}\label{eq:muebetan}
\mu_e^{\beta n}(A,Z)\equiv \frac{M(A-\Delta N,Z-\Delta Z)c^2-M(A,Z)c^2 +m_n c^2 \Delta N}{\Delta Z} + m_e c^2 \, ,
\end{equation}
and we have assumed $\Delta Z>0$. We have also neglected the term arising from the expansion of the neutron chemical potential 
since 
\begin{equation}
 \frac{\Delta N}{Z-\Delta Z} \frac{d\mu_n/dn_n}{d\mu_e/dn_e} =  \frac{dP_n/dn_n}{dP_e/dn_e} \ll 1\, .
\end{equation}
In case of neutron emission without any electron capture (i.e. $\Delta Z=0$ and $\Delta N>0$), we
find
\begin{equation}\label{eq:cond-n-emission-mean}
M(A,Z)- M(A-\Delta N, Z)  < \Delta N \frac{\mu_n}{c^2} \, .
\end{equation}
It is to be understood that $\mu_n$ and its derivative are evaluated at density $n_e \Delta N /( Z-\Delta Z)$. Therefore, 
$\mu_n>m_n c^2$ whenever $\Delta N>0$. In the limiting case $\Delta Z=Z$, the stability condition takes a form similar 
to Eq.(\ref{eq:n-drip-mue}) except that $m_n c^2$ in Eq.(\ref{eq:muedrip}) is now replaced by $\mu_n(n^+)> m_n c^2$. 

This analysis thus leads to a discontinuous change of the neutron density hence also of the neutron chemical potential 
(unless of course $\Delta N=0$). Moreover, as can be seen by comparing Eq.~(\ref{eq:cond-n-emission-mean}) to 
Eq.~(\ref{eq:cond-n-emission}), the nuclei $^A_ZX$ are not necessarily stable against neutron emission contrary 
to the general considerations of Section~\ref{misconceptions}. 
These unphysical results stem mainly from our implicit assumption (arising from the mean-nucleus approximation) 
that all nuclei in a given crustal layer (i.e. at a given pressure $P$) will emit neutrons. 
In reality, only some fraction of nuclei may become unstable against neutron emission so that the crust will be most presumably
composed of an admixture of nuclei $^A_ZX$ and $^{A-\Delta N}_{Z-\Delta Z}Y$. 

\subsection{Beyond the mean-nucleus approximation}

Let us consider that after the onset of electron capture and neutron emission processes~(\ref{eq:e-capture+n-emission}),
only some fraction of nuclei $^A_ZX$ will transform into $^{A-\Delta N}_{Z-\Delta Z}Y$.
In this case, the Gibbs free energy per nucleon can be written as
\begin{eqnarray}\label{eq:gibbs-n-mix-def}
g=\frac{n_X}{n}M(A,Z)c^2 + \frac{n_Y}{n} M(A-\Delta N,Z-\Delta Z)c^2+\frac{n_e}{n}\biggl[\mu_e  -m_e c^2 \biggr] + g_L
+ \frac{n_n}{n}\mu_n \, ,
\end{eqnarray}
where $g_L$ denotes the lattice contribution, for which we use for simplicity the linear mixing rule (see, e.g., Section 2.4.7 
in Ref.~\cite{hae07}). The lattice energy density is thus given by 
\begin{equation}
\mathcal{E}^m_L = \frac{C e^2 n_e^{4/3}}{Z-y\Delta Z}\biggl[(1-y) Z^{5/3} + y (Z-\Delta Z)^{5/3}\biggr]\, ,
\end{equation}
where $y=n_Y/(n_X+n_Y)$. For the lattice contribution to the pressure, we find $P_L^m=\mathcal{E}^m_L/3$. 
The lattice term $g_L$ can finally be expressed as
\begin{equation}
 g_L = \frac{\mathcal{E}_L^m+P_L^m}{n}= \frac{4}{3}\frac{\mathcal{E}_L^m}{n}  \, .
\end{equation}
The baryon number density is given by
\begin{equation}
 n= A n_X + (A-\Delta N) n_Y + n_n\, .
\end{equation}
Free neutrons are associated with nuclei $^{A-\Delta N}_{Z-\Delta Z}Y$, therefore
\begin{equation}
n_n = \Delta N n_Y\, .
\end{equation}
The baryon density can thus be expressed as
\begin{equation}
 n= A (n_X + n_Y)\, .
\end{equation}
The electric charge neutrality requires
\begin{equation}
 n_e= Z n_X + (Z-\Delta Z) n_Y\, .
\end{equation}
Using these equations, the electron and neutron densities can be equivalently written as
\begin{equation}
n_e =\frac{Z- y\Delta Z}{A} n\, .
\end{equation}
\begin{equation}
n_n =\frac{y \Delta N}{A} n = \frac{y \Delta N}{Z- y\Delta Z} n_e\, .
\end{equation}
Finally, we obtain for the Gibbs free energy per nucleon
\begin{eqnarray}\label{eq:gibbs-n-mix}
g(A,Z,\Delta N, \Delta Z, n_e,y)=&&(1-y)\frac{M(A,Z)c^2}{A} + y \frac{M(A-\Delta N,Z-\Delta Z)c^2}{A} \nonumber \\
&+&\frac{Z- y\Delta Z }{A}\biggl[\mu_e  -m_e c^2 \biggr] + \frac{\Delta N}{A} y \mu_n \nonumber \\
 &+& \frac{4}{3 A}C e^2 n_e^{1/3} \biggl[(1-y) Z^{5/3} + y (Z-\Delta Z)^{5/3}\biggr]  \, .
\end{eqnarray}
Note that $y$ can be treated as a continuous variable, which measures the degree of
neutronization of matter. Before the onset of electron capture and neutron emission,
$y=0$ so that $n_n=0$ and
\begin{equation}
n_e^- =\frac{Z}{A} n^-\, .
\end{equation}
The pressure is given by Eq.~(\ref{eq:e-capture-P-}), and the Gibbs free energy per nucleon
is given by $g(A,Z,\Delta N, \Delta Z, n_e^-,y=0)$. With increasing pressure, some infinitesimally
small fraction $dy$ of nuclei may become unstable against neutron emission leading to an \emph{infinitesimally
small} neutron density
\begin{equation}\label{eq:dnn}
dn_n=dy \frac{\Delta N}{Z} n_e^+\, .
\end{equation}
This shows that the neutron density varies continuously across the transition contrary to the mean-nucleus
approach discussed in the previous section. For the nucleus $^A_ZX$ to be stable against electron captures and
neutron emissions, the following condition must be satisfied:
\begin{equation}\label{eq:n-drip-mix-condition1}
g(A, Z, \Delta N, \Delta Z, n_e^+, d y) > g(A,Z,\Delta N, \Delta Z,n_e^-, y=0)\, ,
\end{equation}
where the electron densities $n_e^-$ and $n_e^+$ are related by
\begin{equation}\label{eq:e-capture-n-P}
P=P_e(n_e^-)+P_L(n_e^-,Z)=P_e(n_e^+)+P_L^m(n_e^+,Z,\Delta Z,dy) + P_n(dn_n)\, .
\end{equation}
This equation is similar to Eq.(\ref{eq:e-capture-n-P-mean-nucleus}) obtained in the mean-nucleus approximation, except that
the neutron pressure term is now vanishingly small and can thus be dropped. 
Indeed, using Eq.~(\ref{eq:dnn}) we have 
\begin{equation}
 P_n(dn_n) = P_n\left(dy \frac{\Delta N}{Z} n_e^+\right) = \frac{dP_n}{dn_n} dy \frac{\Delta N}{Z} n_e^+\, ,
\end{equation}
where $dP_n/dn_n$ is evaluated at $n_n=0$. 
On the other hand, the many-body theory for a dilute neutron gas with scattering length $a<0$ yields~\cite{fetwa03} 
\begin{equation}
\frac{\mathcal{E}_n}{n_n} \approx m_n c^2+ \frac{3}{5}\frac{\hbar^2 k_\textrm{F}^2}{2 m_n} - 
\frac{\hbar^2 \pi|a| n_n}{m_n}\biggl[1-\frac{6}{35\pi}(11-2 \ln 2) k_\textrm{F} |a|\biggr]\, ,
\end{equation}
so that $dP_n/dn_n \rightarrow 0$ in the limit $n_n=0$. Therefore, $P_n(dn_n)=0$. 

Setting $n_e^+\equiv n_e + \delta n_e$ with $n_e\equiv n_e^-$ in Eq.~(\ref{eq:e-capture-n-P}), we find to first order 
in $\delta n_e$
\begin{equation}\label{eq:deltane+beyond}
 \delta n_e = \left[P_L(n_e,Z)-P_L^m(n_e,Z,\Delta Z,dy)\right] \left(\frac{d P_e}{d n_e}\right)^{-1}\, .
\end{equation}
Expanding Eq.~(\ref{eq:n-drip-mix-condition1}) to first order in $\alpha$ using Eq.~(\ref{eq:gibbs-n-mix}) and (\ref{eq:deltane+beyond}), 
we finally obtain after some algebra
\begin{eqnarray}\label{eq:e-capture+n-emission-gibbs-approx}
\mu_e + C e^2 n_e^{1/3}\biggl[\frac{Z^{5/3}-(Z-\Delta Z)^{5/3}}{\Delta Z} + \frac{1}{3} Z^{2/3}\biggr] <  \mu_e^{\beta n} \, ,
\end{eqnarray}
where
\begin{equation}\label{eq:muebetan2}
\mu_e^{\beta n}(A,Z)\equiv \frac{M(A-\Delta N,Z-\Delta Z)c^2-M(A,Z)c^2 +m_n c^2 \Delta N}{\Delta Z} + m_e c^2 \, ,
\end{equation}
and we have assumed $\Delta Z>0$. In case of neutron emission without any electron capture (i.e. $\Delta Z=0$ and $\Delta N>0$), we
find
\begin{equation}
M(A,Z)- M(A-\Delta N, Z)  < \Delta N m_n \, .
\end{equation}
Note that this last inequality is independent of the electron background, and coincides with the stability condition (\ref{eq:cond-n-emission}). 
In other words, a nucleus unstable against neutron emission in vacuum will be also unstable in
dense matter, and therefore such a nucleus cannot exist in equilibrium. This means that the neutron-drip transition in dense matter
must be triggered by electron captures.

\section{Matter neutronization in neutron stars}
\label{applications}

\subsection{Nonaccreting neutron stars}
We consider here solitary neutron stars formed in supernova explosions of single massive stars (with a mass 
$M\gtrsim 8 M_\odot$, $M_\odot$ being the mass of the Sun). Initially, the newly born neutron star  
is very hot ($T\sim 10^{10}~$K) and fully fluid. This is the ``hot scenario" of crust formation. A few days after the 
neutron star birth the outer layers are still so hot ($T>2\times 10^9~$K, see e.g. Fig. 34 in Ref.\cite{yak01}) that nuclear 
reaction rates are sufficiently high to keep matter close to the nuclear equilibrium corresponding to the minimum of 
the Gibbs free energy per nucleon $g$ at given temperature $T$ (decreasing) and pressure $P$ (increasing). The star 
cools by neutrino emission and after a few months its outer layer crystallizes forming a solid crust beneath an ocean 
of a fluid hot plasma. Following Refs.~\cite{hw58,htww65}, it is 
assumed that during further cooling the crust remains in nuclear equilibrium until it eventually becomes cold and fully 
``catalyzed''. In order to reach this state, all possible kinds of electroweak and nuclear reactions should have sufficient 
time to be completed. This means that electron captures and neutron emission processes (\ref{eq:e-capture+n-emission}) 
with all possible values of $\Delta Z$ and $\Delta N$ should be considered. The neutron drip transition can thus be found by
minimizing Eq.~(\ref{eq:muebetan}) with respect to $\Delta Z$ and $\Delta N$. Obviously, the number $\Delta Z$ of electrons 
that can be captured by a nucleus $^A_ZX$ must be lower than $Z$, and the number $\Delta N$ of neutrons that can be 
emitted must be lower than $A$. Moreover, the number of bound neutrons in the daughter nucleus $^{A-\Delta N}_{Z-\Delta Z}Y$ 
should be positive, thus leading to $\Delta Z\geq Z+\Delta N - A$. On the other hand, we have previously shown that the neutron 
emission must be accompanied by electron captures so that $\Delta Z>0$ therefore $\Delta N >  A - Z$. All in all, we have
\begin{equation}
Z \geq \Delta Z > Z+\Delta N - A \, , \hskip0.5cm   A \geq \Delta N > A - Z \, .
\end{equation}
Using Eqs.~(\ref{eq:mue-ultra-rel}),  (\ref{eq:gibbs-n-mix}), (\ref{eq:e-capture+n-emission-gibbs-approx}), and (\ref{eq:muebetan2}), 
the Gibbs free energy per nucleon at the neutron drip transition is given to lowest order in $\alpha$ by
\begin{eqnarray}
g(A,Z,\Delta N, \Delta Z, n_e,y=0)&&=\frac{M(A,Z)c^2}{A}\left(1-\frac{Z}{\Delta Z}\right) + \frac{Z}{\Delta Z}\frac{M(A-\Delta N,Z-\Delta Z)c^2}{A} \nonumber \\
&+&  \frac{C \alpha \mu_e^{\beta n}}{A(3 \pi^2)^{1/3}}  \biggl[ Z^{5/3}\left(1-\frac{Z}{\Delta Z}\right) 
+\frac{Z (Z-\Delta Z)^{5/3}}{\Delta Z}\biggr] \nonumber \\
&+&\frac{Z \Delta N}{A \Delta Z } m_n c^2 \, .
\end{eqnarray}
The lowest value of the Gibbs free energy per nucleon, $g=m_n c^2$, is thus obtained for $\Delta Z=Z$ and $\Delta N=A$: the electron captures
proceed until the complete disintegration of the nuclei. Similar considerations were previously discussed in the seminal work of Ref.~\cite{bps}. 
It can be easily seen that in this case Eq.~(\ref{eq:e-capture+n-emission-gibbs-approx}) reduces to Eq.~(\ref{eq:n-drip-mue}). 
Considering that electrons are ultrarelativistic using Eqs.~(\ref{eq:mue-ultra-rel}) and (\ref{eq:P-ultra-rel}), the baryon density and 
pressure at neutron drip are approximately given by
\begin{equation}\label{eq:rhodrip-multi}
n_{\rm drip}(A,Z) \approx \frac{A}{Z} \frac{\mu_e^{\rm drip}(A,Z)^3}{3\pi^2 (\hbar c)^3}
\biggl[1+\frac{4 C \alpha}{(81\pi^2)^{1/3}} Z^{2/3} \biggr]^{-3}\, ,
\end{equation}
\begin{equation}\label{eq:Pdrip-multi}
P_\textrm{drip}(A,Z) \approx \frac{\mu_e^\textrm{drip}(A,Z)^4}{12 \pi^2 (\hbar c)^3}\biggl[1+\frac{4C \alpha Z^{2/3}}{(81\pi^2)^{1/3}} \biggr]^{-3} \, ,
\end{equation}
with $\mu_e^{\rm drip}(A,Z)$ given by Eq.(\ref{eq:muedrip}). 

\subsection{Accreting neutron stars}

The accretion of matter onto a neutron star from a stellar companion may change the initial constitution of its crust, which was 
originally formed in the ``hot scenario" described in the previous section. For an accretion rate $\dot{M}=10^{-9}~{\rm M_\odot}$/yr 
the original outer crust is replaced by accreted matter in $10^4\;$yr, while the accretion stage in the low-mass binary systems can 
last for $10^9$\;yr. At  densities above $\sim 10^8$~g~cm$^{-3}$ matter is highly degenerate and relatively cold  ($T\lesssim 5\times
 10^8$~K) so that thermonuclear processes are strongly suppressed; their rates are many orders of magnitude lower than the compression
 rate due to accretion. This is the ``cold scenario" of crust formation. On the other hand, the matter composition can still be altered due 
 to electron captures, neutron emissions, and at high enough densities pycnonuclear reactions. Multiple electron captures are unlikely
 and need not be considered.  For example, the double electron capture $2e^-+{\rm ^{56}Fe}\longrightarrow {\rm ^{56}Mn}+2\nu_e$ 
 occurs on a timescale of $10^{20}\;$yr \cite{blaes1990}. 
The onset of neutron drip can thus be determined from Eqs.~(\ref{eq:e-capture+n-emission-gibbs-approx}) and (\ref{eq:muebetan2}) with
$\Delta Z=1$ and $\Delta N>0$. In the limit of ultra relativistic electrons, using Eq.~(\ref{eq:mue-ultra-rel}), the average baryon 
density for the onset of electron captures is approximately given by
\begin{equation}\label{eq:rhodrip-acc}
 n_\textrm{drip-acc}(A,Z) \approx \frac{A}{Z} \frac{\mu_e^\textrm{drip-acc}(A,Z)^3}{3\pi^2 (\hbar c)^3} 
  \biggl[1+\frac{C \alpha}{(3\pi^2)^{1/3}}\left(Z^{5/3}-(Z-1)^{5/3}+\frac{Z^{2/3}}{3}\right)\biggr]^{-3}\, ,
\end{equation}
\begin{equation}\label{eq:muedrip-acc}
\mu_e^\textrm{drip-acc}(A,Z)= M(A-\Delta N,Z-1)c^2-M(A,Z)c^2+\Delta N m_n c^2 + m_e c^2 \, .
\end{equation}
The threshold pressure $P_\textrm{drip-acc}$ can be obtained from Eq.~(\ref{eq:P-ultra-rel}) with $\mu_e$ obtained from Eq.~(\ref{eq:e-capture+n-emission-gibbs-approx}):
\begin{equation}\label{eq:Pdrip-acc}
P_\textrm{drip-acc}(A,Z) \approx \frac{\mu_e^\textrm{drip-acc}(A,Z)^4}{12 \pi^2 (\hbar c)^3}\biggl[1+\frac{4C \alpha Z^{2/3}}{(81\pi^2)^{1/3}} \biggr]
 \biggl[1+\frac{C \alpha}{(3\pi^2)^{1/3}}\left(Z^{5/3}-(Z-1)^{5/3}+\frac{Z^{2/3}}{3}\right)\biggr]^{-4}\, .
\end{equation}

For the sake of comparison, let us calculate the neutron drip
transition in accreting neutron stars  using a strict mean-nucleus
approximation.
For this purpose, we will apply the Mackie-Baym compressible liquid
drop model widely used in previous calculations of accreted
neutron-star crusts~\cite{hz1990a,hz2003,hz2008}.
Assuming that the ashes of the X-ray bursts consist of pure
$^{56}$Fe, the equilibrium nuclei were found to become progressively
more neutron rich with increasing density due to electron captures.
At density $5.65\times 10^{11}$~g~cm$^{-3}$ corresponding to the pressure 
$1.23\times 10^{30}$ dyn cm$^{-2}$, the nucleus $^{56}$Ar
is found to be unstable against one neutron emission,
\begin{equation}
 ^{56}\textrm{Ar} + e^-  \rightarrow ^{55}\textrm{Cl} +  n +\nu_e\, .
 \label{eq:mbdrip56}
\end{equation}
Solving numerically Eq.~(\ref{eq:e-capture+n-emission-gibbs-approx}) with
$\Delta N=1$ and the masses calculated with the same compressible
liquid drop model as in Refs.~\cite{hz1990a,hz2003} in the absence of a neutron 
gas, we now find that the neutron drip transition actually occurs at a lower density,
given by $\rho_\textrm{drip-acc}\simeq 5.09\times 10^{11}$~g~cm$^{-3}$. The neutron-drip 
pressure is now given by $P_\textrm{drip-acc}\simeq 1.07\times 10^{30}$ dyn cm$^{-2}$. 
For ashes made of $^{106}$Pd, using the Mackie-Baym compressible liquid
drop model of accreted crust, the neutron drip nucleus $^{106}$Ge is
found to decay into $^{103}$Ga,
\begin{equation}\label{eq:mbdrip106}
 ^{106}\textrm{Ge} + e^-  \rightarrow ^{103}\textrm{Ga} + 3 n +\nu_e\, ,
\end{equation}
at density $6.66\times 10^{11}$~g~cm$^{-3}$ (pressure $1.38\times 10^{30}$ dyn cm$^{-2}$). 
Solving now numerically Eq.~(\ref{eq:e-capture+n-emission-gibbs-approx}) with $\Delta N=3$, the neutron-drip 
density decreases to $\rho_\textrm{drip-acc}\simeq 5.31\times 10^{11}$~g~cm$^{-3}$. 
The drip pressure is $P_\textrm{drip-acc}\simeq 1.02\times 10^{30}$ dyn cm$^{-2}$.
Let us remark that the nuclei $^{55}$Cl and $^{103}$Ga are unstable and undergo further 
electron captures accompanied by neutron emissions. 
We found that the errors of the analytical expressions (\ref{eq:rhodrip-acc}) and 
(\ref{eq:Pdrip-acc}) lie below about $0.1\%$. Approximating the neutron drip density by 
$\rho_\textrm{drip-acc} \approx m n_\textrm{drip-acc}$ where $m$ is the unified atomic mass unit, and 
using Eq.~(\ref{eq:rhodrip-acc}), leads to an error of about $0.7\%$. 

It should be remarked that our values of the neutron-drip density calculated within
the mean-nucleus approximation differ from those previously obtained in Refs.~\cite{hz1990a,hz2003} because
of a different treatment of neutrons produced by electron captures. In Refs.~\cite{hz1990a,hz2003}, these
neutrons were kept inside the nucleus, assuming that free neutrons outside remained unaffected. The daughter
nucleus was then found to be unstable against neutron emission alone. In the present approach, we minimize
the Gibbs free energy per nucleon (for a given atomic number) without any further constraint. This
new procedure yields values for the neutron-drip density and pressure that are closer to the exact results obtained 
from Eq.~(\ref{eq:e-capture+n-emission-gibbs-approx}). Indeed, using the new version of the Mackie and Baym model of accreted crust for the $^{56}$Fe ashes, 
we are now getting $5.7\times 10^{11}$~g~cm$^{-3}$ instead of $6.1\times 10^{11}$~g~cm$^{-3}$ obtained in Refs.~\cite{hz1990a,hz2003}.
In the case of the $^{106}$Pd ashes the difference is even larger, $6.7\times 10^{11}$~g~cm$^{-3}$ instead of
$7.8\times 10^{11}$~g~cm$^{-3}$ obtained in Refs.~\cite{hz1990a,hz2003}.

%

%

As the nuclei from the ashes of X-ray bursts sink into the crust, their proton number $Z$ decreases due to electron captures whereas
$A$ remains unchanged (we assume that pycnonuclear reactions may only occur in the inner crust, where nuclei are immersed in a
neutron liquid).
At some point, the daughter nuclei will be so neutron rich that neutrons will be emitted. Therefore, the neutron-drip transition will occur
when the threshold electron chemical potential $\mu_e^{\beta n}$ for neutron emission (i.e. $\Delta N>0$ and $\Delta Z=1$) will become
lower than the  threshold electron chemical potential $\mu_e^\beta$ for electron capture alone. Using  Eqs.(\ref{eq:muebeta}) and
(\ref{eq:muebetan2}), the condition $\mu_e^{\beta n}(A,Z) < \mu_e^\beta(A,Z)$ yields
\begin{equation}
S_{\Delta N n}(A,Z-1)\equiv M(A-\Delta N,Z-1)-M(A,Z-1)+\Delta N m_n < 0\, .
\end{equation}
In other words, the nucleus $^A_ZX$ marking the neutron drip point is such that the nucleus $^A_{Z-1}Y$ is unstable against neutron
emission. Consequently, for any given value of the mass number $A$, the proton number $Z$ of the nuclei present in the 
outer crust will decrease until the $\Delta N$-neutron separation energy $S_{\Delta N n}(A,Z-1)$ becomes negative: this will mark
 the onset of neutron drip. 

We have determined in this way the neutron-dripping nucleus in the crust of accreting neutron stars. As for the initial composition of the ashes,
we considered two different scenarios: first, the ashes are produced by an $rp$-process during an X-ray burst~\cite{schatz2001}, and second, the
ashes are produced by steady state hydrogen and helium burning~\cite{schatz2003} as expected to occur during superbursts~\cite{gupta2007}. We have 
calculated the neutron-drip density and pressure by solving numerically Eq.~(\ref{eq:e-capture+n-emission-gibbs-approx}) without any further approximation
considering  all possible neutron emission processes. 
Results are shown in Tables~\ref{tab2}, \ref{tab3} and \ref{tab4} for the three different microscopic nuclear mass models HFB-19, HFB-20 and HFB-21, which are
based on the self-consistent Hartree-Fock-Bogoliubov method~\cite{gcp10}. The neutron drip density and pressure in accreting neutron stars are thus found to
be quite sensitive to the composition of the ashes. In reality, the crust of an accreting neutron star is expected to contain an admixtures of various nuclides.
Therefore, the onset of neutron drip will be determined by the most unstable nucleus. Depending on the adopted mass model, the threshold density is thus found to 
vary from $2.80\times 10^{11}$ g~cm$^{-3}$ to $6.13\times 10^{11}$ g~cm$^{-3}$ ($P_{\rm drip-acc}$ ranging from $4.72\times 10^{29}$ dyn~cm$^{-2}$ to 
$13.1\times 10^{29}$ dyn~cm$^{-2}$) for ordinary burst ashes, and from $3.23\times 10^{11}$ g~cm$^{-3}$ to $6.13\times 10^{11}$ g~cm$^{-3}$ ($P_{\rm drip-acc}$
ranging from $6.11\times 10^{29}$ dyn~cm$^{-2}$ to $13.1\times 10^{29}$ dyn~cm$^{-2}$) for superburst ashes. For comparison, as indicated in Table~\ref{tab1},
the neutron-drip density in catalyzed matter lies in the range $4.30-4.40\times 10^{11}$ g~cm$^{-3}$ ($P_{\rm drip}$ ranging from $7.84\times 10^{29}$ dyn~cm$^{-2}$ to
$7.91\times 10^{29}$ dyn~cm$^{-2}$). This analysis shows that the neutron-drip transition in accreting neutron stars may occur either at a higher or at a lower
density than in nonaccreting neutron stars.
In tables~\ref{tab5} and \ref{tab6}, the predictions from the microscopic nuclear mass models are compared to those from the Mackie and Baym compressible liquid drop model
employed in Refs.~\cite{hz1990a,hz2003}. For this purpose, we have considered the same initial composition of ashes as in Refs.~\cite{hz1990a,hz2003}.
For $^{56}$Fe ashes, all models predict the same atomic number $Z=18$ for the dripping nucleus and the same number $\Delta N=1$ of emitted neutrons. On the other hand,
the HFB models predict substantially lower values for the average neutron drip density and pressure. Much larger deviations are found in the case of $^{106}$Pd ashes, mainly
due to different predictions for the dripping nucleus and the number of emitted neutrons. In particular, the neutron-drip densities and pressures, as obtained from
HFB-21 (which is favored over the two other HFB models for the reasons given in Refs.~\cite{cha11,fant13,ho15}) and the compressible liquid drop model, differ by almost a factor 
a two. This suggests that the composition of accreted crusts could be substantially different from those found in Refs.~\cite{hz1990a,hz2003,hz2008}.

\begin{table}
\centering
\caption{Neutron drip transition in the crust of accreting neutron stars, as predicted by the HFB-19 microscopic nuclear mass
model: mass and atomic numbers of the dripping nucleus, number of emitted neutrons, density and corresponding pressure.
The mass numbers $A$ are listed from top to bottom considering that the ashes are produced by ordinary
X-ray bursts (upper panel) or superbursts (lower panel). See text for details.}\smallskip
\label{tab2}
\begin{tabular}{ccccc}
\hline 
  A   &  Z & $\Delta N$ &  $\rho_{\rm drip-acc}$ ($10^{11}$ g~cm$^{-3}$)  & $P_{\rm drip-acc}$ ($10^{29}$ dyn~cm$^{-2}$) \\
 \hline 
 104  &  32   &   1  &   4.76  &  9.08  \\
 105  &  33   &   1  &   3.31  &  5.74  \\
  68  &  22   &   1  &   4.00  &  7.78  \\
  64  &  20   &   3  &   5.74  &  12.0  \\
  72  &  22   &   1  &   5.15  &  10.1  \\
  76  &  24   &   1  &   4.97  &  10.0  \\
  98  &  32   &   1  &   3.34  &  6.13  \\
 103  &  33   &   1  &   2.82  &  4.76  \\
 106  &  34   &   1  &   3.65  &  6.73  \\
   \hline
  66  &  22   &   1  &   3.46  &  6.68  \\
  64  &  20   &   3  &   5.74  &  12.0  \\
  60  &  20   &   1  &   3.23  &  6.11  \\
\end{tabular}
\end{table}

\begin{table}
\centering
\caption{Same as Table~\ref{tab2} for the HFB-20 microscopic nuclear mass model.}\smallskip
\label{tab3}
\begin{tabular}{ccccc}
\hline 
  A   &  Z  & $\Delta N$&  $\rho_{\rm drip-acc}$ ($10^{11}$ g~cm$^{-3}$)  & $P_{\rm drip-acc}$ ($10^{29}$ dyn~cm$^{-2}$)\\
 \hline 
104  & 32  &  1  &  4.82  &  9.23  \\
105  & 33  &  1  &  3.38  &  5.91  \\
 68  & 22  &  1  &  4.09  &  8.01  \\
 64  & 20  &  3  &  6.13  &  13.1  \\
 72  & 22  &  1  &  5.21  &  10.3  \\
 76  & 24  &  1  &  5.01  &  10.1  \\
 98  & 32  &  1  &  3.34  &  6.14  \\
103  & 33  &  1  &  2.80  &  4.72  \\
106  & 32  &  1  &  5.26  &  10.1  \\
       \hline
 66  & 22  &  1  &  3.54 &  6.88   \\
 64  & 20  &  3  &  6.13 &  13.1   \\
 60  & 20  &  1  &  3.27 &  6.20   \\
 \end{tabular}
 \end{table}
 
 \begin{table}
\centering
\caption{Same as Table~\ref{tab2} for the HFB-21 microscopic nuclear mass model.}\smallskip
\label{tab4}
\begin{tabular}{ccccc}
\hline 
  A   &  Z  & $\Delta N$&  $\rho_{\rm drip-acc}$ ($10^{11}$ g~cm$^{-3}$)  & $P_{\rm drip-acc}$ ($10^{29}$ dyn~cm$^{-2}$)\\
 \hline 
 104  &   32 &  1 &   4.85 &   9.31  \\
 105  &   33 &  1 &   3.42 &   6.01  \\
  68  &   22 &  1 &   4.13 &   8.12  \\
  64  &   20 &  3 &   5.84 &   12.3  \\
  72  &   22 &  1 &   5.35 &   10.6  \\
  76  &   24 &  1 &   5.02 &   10.2  \\
  98  &   32 &  1 &   3.42 &   6.33  \\
 103  &   33 &  1 &   2.83 &   4.79  \\
 106  &   34 &  1 &   3.65 &   6.72  \\
  \hline
  66  &   22 &  1 &   3.58 &   6.98  \\
  64  &   20 &  3 &   5.84 &   12.3  \\
  60  &   20 &  1 &   3.36 &   6.43  \\
 \end{tabular}
 \end{table}

\begin{table}
\centering
\caption{Neutron drip transition in the crust of accreting neutron stars, as predicted by different nuclear mass models for
$^{56}$Fe ashes: atomic number $Z$ of the dripping nucleus, number of emitted neutrons, density and corresponding pressure. 
Three different microscopic Hartree-Fock-Bogoliubov nuclear mass models (HFB) are compared
to the compressible liquid drop model of Mackie and Baym (MB). See text for details.}\smallskip
\label{tab5}
\begin{tabular}{ccccc}
\hline 
                                              & HFB-19 & HFB-20  & HFB-21 &  MB  \\
\hline 
$Z$                                           &   18   &  18     &  18    &  18  \\
$\Delta N$                                    &    1   &   1     &   1    &  1   \\
$\rho_{\rm drip-acc}$ ($10^{11}$ g~cm$^{-3}$) & 4.49   &  4.50   &  4.38  &  5.09 \\
$P_{\rm drip-acc}$ ($10^{29}$ dyn~cm$^{-2}$)  & 9.02   &  9.06   &  8.74  &  10.7 \\
\end{tabular}
\end{table}

\begin{table}
\centering
\caption{Same as Table~\ref{tab5} for $^{106}$Pd ashes. }\smallskip
\label{tab6}
\begin{tabular}{ccccc}
\hline 
                                              & HFB-19 & HFB-20  & HFB-21 &  MB  \\
\hline 
$Z$                                           &   34   &  32     &  34    &  32  \\
$\Delta N$                                    &    1   &   1     &   1    &  3   \\
$\rho_{\rm drip-acc}$ ($10^{11}$ g~cm$^{-3}$) & 3.65   &  5.26   &  3.65  &  5.31 \\
$P_{\rm drip-acc}$ ($10^{29}$ dyn~cm$^{-2}$)  & 6.73   &  10.1   &  6.72  &  10.2  \\
\end{tabular}
\end{table}

\section{Conclusions}

With increasing pressure, dense matter in the outer crust of a neutron star becomes progressively more neutron rich. At some point,
neutrons start to drip out of nuclei. In both accreting and nonaccreting neutron stars, the nucleus at the onset of neutron drip is shown to 
be stable against neutron emission, but is actually unstable against electron capture accompanied by neutron emission. After examining the 
occurrence of such processes in both accreting and nonaccreting neutron star crusts, we have obtained general analytical expressions for the 
density and pressure at the onset of neutron drip considering that only a vanishingly small fraction of nuclei become unstable. In this way, 
we have shown that the spurious discontinuous change in the density of unbound neutrons that was found in
previous studies of accreting neutron stars~\cite{hz1990a,hz2003} arises from the use of the mean-nucleus approximation. As a
consequence, this approximation overestimates the neutron-drip density and pressure.

We have also studied numerically the transition between
the outer and inner crusts of accreting neutron stars for various initial compositions of the ashes. In particular, we considered ashes produced from
both X-ray bursts and superbursts. For this purpose, we have made use of experimental atomic mass data complemented with microscopic atomic
mass tables. The neutron-drip density is found to be shifted to either lower or higher values than in nonaccreting neutron stars, depending on whether accreting
neutron stars exhibit ordinary bursts or superbursts. The crust of accreting neutron stars is also predicted to contain various ultradrip nuclei so that it
is necessary to extend the calculations of atomic masses beyond the neutron drip line.
Finally, large deviations were found between the predictions of microscopic mass models and the more phenomenological liquid drop models employed in
Refs.~\cite{hz1990a,hz2003}, thus suggesting that the composition and the properties of accreted neutron stars crusts could differ substantially from those
predicted in Refs.~\cite{hz1990a,hz2003}. 

\appendix
\section{Neutron drip transition in cold catalyzed matter with the Harrison-Wheeler model}
\label{appendix}

In the seminal work from Wheeler and collaborators~\cite{hw58,htww65} (see also Ref.~\cite{shapiro1983}), nuclear masses were calculated
using the semi-empirical formula of Green~\cite{green55}
\begin{equation}\label{eq:green}
M(A,Z)= m_u \biggl[ b_1 A + b_2 A^{2/3} - b_3 Z + b_4 A \left(\frac{1}{2} - \frac{Z}{A}\right)^2 + b_5 \frac{Z^2}{A^{1/3}}\biggr] \, ,
\end{equation}
where $b_1 = 0.992064$, $b_2 = 0.01912$, $b_3 = 0.00084$, $b_4 = 0.10178$, $b_5 = 0.000763$, and $m_u\equiv M(^{16}\textrm{O})/16$.
The equilibrium nucleus present in the crust of a nonaccreting neutron star can be determined by minimizing the Gibbs free energy per
nucleon $g$ with respect to both $Z$ and $A$ for any given pressure $P$. Neglecting the lattice contribution to $g$ and treating $Z$
and $A$ as continuous variables lead in particular to the following relation~\cite{htww65}
\begin{equation}\label{eq:ZvsA}
Z=\left(\frac{b_2}{2 b_5}\right)^{1/2} A^{1/2}\, .
\end{equation}
The actual values of $Z$ and $A$ can be found from the beta-equilibrium condition~\cite{htww65}
\begin{equation}
\frac{\partial M(A,Z)}{\partial Z} = m_e -\mu_e/c^2 \, ,
\end{equation}
where the free electron chemical potential is obtained from the requirement that the electron pressure be given by $P$.
The neutron-drip transition occurs at some pressure $P_\textrm{drip}$ such that the following condition is fulfilled:
\begin{equation}\label{eq:ndrip-continuum}
\frac{\partial M(A,Z)}{\partial A} = m_n \, .
\end{equation}
Solving Eq.~(\ref{eq:ndrip-continuum}) using Eqs.~(\ref{eq:ZvsA}) and (\ref{eq:green}) yields $Z_\textrm{drip}\simeq 39.09$ and
$A_\textrm{drip}\simeq 122.0$. We have adopted the same value for the neutron mass as in Ref.~\cite{htww65}, namely $m_n=1.008982 m_u$.
The dripping nucleus is stable against neutron emission:
\begin{equation}
M(A_\textrm{drip},Z_\textrm{drip}) - M(A_\textrm{drip}-1,Z_\textrm{drip}) - m_n \simeq -8.6\times 10^{-5} m_u < 0 \, .
\end{equation}

 \begin{acknowledgments}
This work was financially supported by FNRS (Belgium), the Polish National Science Centre
through the OPUS grant 2013/11/ST9/04528,  the Simons Foundation (USA) and the COST Action MP1304.
This work was initiated at the Aspen Center for Physics (Colorado, USA), and was further discussed at the International Space
Science Institute located in Bern (Switzerland). The hospitality of these two institutions is gratefully acknowledged.
\end{acknowledgments}


\begin{thebibliography}{99}
\bibitem{hae07} P. Haensel, A.~Y. Potekhin, and D.~G. Yakovlev, \textit{Neutron Stars 1: Equation of state
and structure}, Springer (2007).
\bibitem{lrr} N. Chamel and P. Haensel,``Physics of Neutron Star Crusts'',
Living Rev. Relativity 11, (2008), 10. http://www.livingreviews.org/lrr-2008-10
\bibitem{bps} G. Baym, C. Pethick, and P. Sutherland, Ap. J. {\bf 170}, 299 (1971).
\bibitem{pearson2011} J.~M. Pearson, S. Goriely, N. Chamel, Phys.~Rev.~C {\bf 83}, 065810 (2011).
\bibitem{wolf13} R. N. Wolf et al., Phys. Rev. Lett. {\bf 110}, 041101 (2013).
\bibitem{hemp13} S. Kreim, M. Hempel, D. Lunney, and J. Schaffner-Bielich, Int. J. Mass Spec. {\bf 349-350}, 63 (2013).
\bibitem{hz1990a} P. Haensel, J.-L. Zdunik, A\&A {\bf 227}, 431 (1990).
\bibitem{hz2003} P. Haensel, J.-L. Zdunik, A\&A {\bf 404}, L33 (2003).
\bibitem{baiko01} Baiko et al., Phys. Rev. E {\bf 64}, 057402 (2001).
\bibitem{hw58} B.~K. Harrison, and J.~A. Wheeler, in \textit{Onzi\`eme Conseil de Physique Solvay}, Stoops, Bruxelles, Belgium (1958).
\bibitem{htww65} B.~K. Harrison, K.~S. Thorne, M. Wakano, and J.~A. Wheeler, \textit{Gravitation Theory and Gravitational Collapse},
The University of Chicago Press (1965).
\bibitem{cha06} N. Chamel, Nucl. Phys. {\bf A773}, 263 (2006).
\bibitem{cha07}N. Chamel, S. Naimi, E. Khan and J. Margueron, Phys. Rev. C {\bf 75}, 055806 (2007).
\bibitem{cch05} B. Carter, N. Chamel and P. Haensel, Nucl. Phys. {\bf A 748}, 675 (2005).
\bibitem{hz1989} P. Haensel, J.-L. Zdunik, J. Dobaczewski, A\&A {\bf 222}, 353 (1989).
\bibitem{shapiro1983} S.~L. Shapiro and S.~A. Teukolsky, \textit{Black Holes, White Dwarfs, and Neutron Stars}, John Wiley\&Sons (1983).
\bibitem{fetwa03} A.~L. Fetter and J.~D. Walecka, \textit{Quantum Theory of Many-Particle Systems}, Dover, New York (2003), p.149. 
\bibitem{yak01} D.G. Yakovlev, A.D. Kaminker, O.Y. Gnedin, P. Haensel, Phys. Repts. {\bf 354}, 1 (2001).
\bibitem{blaes1990} O. Blaes, R. Blanford, P. Madau, S. Koonin, ApJ
{\bf 363}, 612 (1990).
\bibitem{hz2008} P. Haensel, J.-L. Zdunik, A\&A {\bf 480}, 459 (2008).
\bibitem{schatz2001} H. Schatz et al. Phys. Rev. Lett.{\bf 86}, 3471 (2001).
\bibitem{schatz2003} H. Schatz, L. Bildsten, A. Cumming, M. Ouellette, Nucl. Phys. {\bf A718}, 247 (2003).
\bibitem{gupta2007} S. Gupta, E. F. Brown, H. Schatz, P. M\"oller, K.-L. Kratz, ApJ {\bf 662}, 1188 (2007).
\bibitem{gcp10}S. Goriely, N. Chamel, and J.~M. Pearson, Phys. Rev. C {\bf 82}, 035804 (2010).
\bibitem{cha11} N. Chamel, A.~F. Fantina, J.~M. Pearson, S. Goriely, Phys. Rev. C {\bf 84}, 062802(R) (2011).
\bibitem{fant13} A.~F. Fantina, N. Chamel, J.~M. Pearson, S. Goriely, A\&A {\bf 559}, A128 (2013). 
\bibitem{ho15} Wynn C. G. Ho, K. G. Elshamouty, C. O. Heinke, and A. Y. Potekhin, Phys. Rev. C {\bf 91}, 015806 (2015)
\bibitem{green55} A. E. S. Green, \textit{Nuclear Physics} (New York, Mc Graw-Hill Book Co., 1955).
\end{thebibliography}
\end{document}